\documentclass[12pt]{article}

\usepackage[T1]{fontenc}
\usepackage[utf8]{inputenc}   % if compiling with pdfLaTeX
\usepackage{lmodern}
\usepackage[polish,english]{babel} % English active (last)

\usepackage{amsmath, amssymb}
\usepackage{graphicx}
\usepackage{float}

\usepackage[hidelinks]{hyperref}

\usepackage{microtype}

\DeclareMathOperator{\artanh}{artanh}

\title{\LARGE \textbf{Comment on Marek Czachor article entitled ``On Relativity of Quantumness as Implied by Relativity of Arithmetic and Probability''}}
\author{
\large Mikołaj Sienicki\thanks{Polish-Japanese Academy of Information Technology, ul. Koszykowa 86, 02-008 Warsaw, Poland, European Union.}
\quad and \quad
Krzysztof Sienicki\thanks{Chair of Theoretical Physics of Naturally Intelligent Systems ($\mathbb{NIS}^{\text{\textcopyright{}}}$), Lipowa 2/Topolowa 19, 05-807 Podkowa Leśna, Poland, European Union.}
}
\date{\today}

\begin{document}
\maketitle

\begin{abstract}
Czachor’s model of hierarchical arithmetics begins with a valid formal premise but fixes the key probability mapping $g$ by importing the Born rule and Fubini--Study metric from standard quantum mechanics, where Born probabilities are Kolmogorov \emph{within a fixed measurement context}. 
This $g$ is then applied in a non-Newtonian hidden-variable setting, producing a hybrid framework whose agreement with quantum correlations is built in by design, not derived from new physics, and thus does not constitute a genuine counterexample to Bell’s theorem. Moreover, the construction \emph{changes a standard premise} used in one precise formulation of Bell’s theorem (classical Kolmogorov probability), and therefore lies outside that theorem’s original scope.
\end{abstract}

The starting point of \cite{Czachor_Hierarchy} is both elegant and correct: if you change the underlying arithmetic, you change how probabilities combine. A whole \emph{hierarchy} of isomorphic arithmetics thus gives a hierarchy of probabilistic models. Formally, this leaves room for infinitely many bijections \(g\) that translate probabilities between levels. So far, so good.

What follows in the paper, however, hinges on a modeling choice that quietly mixes two different frameworks. Below I explain where that happens and why it matters.

\subsection*{Step 1 — The purely formal setup}
We posit arithmetics \(A_1,A_2,\dots\), each with its own \(\oplus_k,\otimes_k,\ldots\), all isomorphic via \(f_{ij}:A_i\to A_j\).
Because probability theory depends on arithmetic to combine events, each \(A_k\) induces a (potentially different) probabilistic model.

At this abstract stage, no specific \(g\) is fixed; many are possible.

\subsection*{Step 2 — Connecting hidden and observable levels}
Pick two layers: a hidden-variable level \(A_h\) (HV) and an observable, quantum-mechanical level \(A_q\) (QM). We relate their probabilities by a bijection
\begin{equation}
P_{\mathrm{QM}} = g(P_{\mathrm{HV}}).
\end{equation}

\textbf{Assumption 1 (Complement preservation).} We explicitly assume the regraduation respects complements,
\[
g(p)+g(1-p)=1.
\]
\emph{Immediate consequence.} By Assumption~1, \(g(\tfrac{1}{2})=1-g(\tfrac{1}{2})\), hence \(g(\tfrac{1}{2})=\tfrac{1}{2}\).

\emph{Admissibility.} To behave like a sensible “regrading” of probabilities, we also assume \(g\) is continuous, strictly increasing on \([0,1]\), and obeys the boundaries \(g(0)=0\), \(g(1)=1\). These conditions preserve event ordering and keep the mapping probabilistically well-behaved. \textit{These properties make $g$ a bijection $[0,1]\!\to\![0,1]$.}

\subsection*{Step 3 — Calibrating the observable layer with standard QM}
On the observable side we use standard quantum mechanics. Born’s rule gives
\begin{equation}
P_{\mathrm{QM}}(b|a)=|\langle b \mid a\rangle|^2.
\end{equation}
For \emph{pure states}, the Fubini–Study (FS) distance is
\begin{equation}
d_{\mathrm{FS}}(a,b)=\arccos|\langle b \mid a\rangle|
\quad\Rightarrow\quad
P_{\mathrm{QM}}(b|a)=\cos^2 d_{\mathrm{FS}}(a,b).
\end{equation}
For qubits it is convenient to use the Bloch-sphere angle \(\theta\), with \(d_{\mathrm{FS}}=\theta/2\), hence
\begin{equation}
P_{\mathrm{QM}}(b|a)=\cos^2\!\Big(\frac{\theta}{2}\Big).
\end{equation}
\noindent Here \(\theta\) is the angle between the corresponding Bloch vectors of the pure states \(|a\rangle\) and \(|b\rangle\).

\noindent\emph{Caveat.} The FS formula above applies to \emph{pure} states. For mixed states a natural geometric replacement uses quantum fidelity (Bures angle) \(\mathcal{B}(\rho,\sigma)=\arccos\!\sqrt{F(\rho,\sigma)}\). However, there is \emph{no universal identity} of the form \(P=\cos^2 \mathcal{B}\) for general mixed states; probabilities remain \(P(b|\rho)=\mathrm{tr}[\rho\,\Pi_b]\) for a specified POVM, so any “calibration” beyond the pure-state case must state the measurement model explicitly.

\textbf{Key idea.} This calibration imports standard quantum structure at the observable level; in particular, within a \emph{fixed measurement context} the Born probabilities are Kolmogorov. The hidden-variable layer, by contrast, is allowed to use a non-Newtonian calculus. The result is inherently hybrid.

\noindent\emph{By a fixed measurement context} we mean a single POVM or, in the projective case, a commuting set of observables (a Boolean subalgebra of events). Within such a context, Born probabilities admit a Kolmogorov description; globally, the quantum event lattice is non-Boolean.

\subsection*{Step 4 — Choosing a specific link \(g\)}
The paper then makes an ansatz that ties the Bloch angle \(\theta\) linearly to the hidden probability \(p\):
\begin{equation}
  \theta \;=\; \pi\,(1-p).
\end{equation}
Plugging this into \(P_{\mathrm{QM}}=\cos^2(\theta/2)\) gives
\begin{equation}
  \begin{aligned}
    P_{\mathrm{QM}}
      &= \cos^2\!\Big(\frac{\pi}{2}\,(1-p)\Big) \\
      &= \sin^2\!\Big(\frac{\pi p}{2}\Big),
  \end{aligned}
\end{equation}
so that
\begin{equation}
  \begin{aligned}
    g(p)
      &= \sin^2\!\Big(\frac{\pi p}{2}\Big) \\
      &= \frac{1-\cos(\pi p)}{2}.
  \end{aligned}
\end{equation}

\emph{Remark.} The linear relation \(\theta=\pi(1-p)\) is an \emph{illustrative modeling choice}, not a consequence of Assumption~1. Many other choices would lead to different \(g\)’s. However, to satisfy Assumption~1 (complement preservation) the parametrization must obey the symmetry
\[
\theta(1-p)=\pi-\theta(p),
\]
since then
\[
g(p)+g(1-p)=\cos^2\!\frac{\theta(p)}{2}+\cos^2\!\frac{\theta(1-p)}{2}
=\cos^2 x+\sin^2 x=1
\quad\text{with }x=\tfrac{\theta(p)}{2}.
\]
Equivalently, one may define \(g\) arbitrarily on $[0,\tfrac12]$ (subject to continuity/monotonicity, $g(0)=0$, \emph{and} \(g(\tfrac12)=\tfrac12\) to ensure continuity at the join) and extend by $g(1-p)=1-g(p)$. \emph{Thus not every monotone $\theta$ yields an admissible $g$; the symmetry (or complement-extension) is required.}

To illustrate non-uniqueness while keeping the basic constraints, here are two distinct, admissible examples:
\begin{equation}
  g_1(p) \;=\; 3p^2 - 2p^3
  \qquad\text{(smooth, monotone)},
\end{equation}
\begin{equation}
  g_{\text{alt}}(p)
  \;=\;
  \sin^{2}\!\Big(
    \frac{\pi}{4}\,\big[\,1+\sin\!\big(\pi\,(p-\tfrac{1}{2})\big)\big]
  \Big)
  \qquad\text{(smooth, monotone)}.
\end{equation}

\noindent\textit{Define} \(s:[0,1]\to[0,1]\) by
\[
s(p)=\tfrac{1}{2}+\tfrac{1}{2}\sin\!\big(\pi(p-\tfrac{1}{2})\big),
\quad\text{so that}\quad
g_{\text{alt}}(p)=\sin^{2}\!\big(\tfrac{\pi}{2}\,s(p)\big).
\]
\textit{Symmetry.} Since \(s(1-p)=1-s(p)\),
\[
g_{\text{alt}}(p)+g_{\text{alt}}(1-p)
=\sin^{2}\!\big(\tfrac{\pi}{2}s(p)\big)+\sin^{2}\!\big(\tfrac{\pi}{2}(1-s(p))\big)=1.
\]
\textit{Monotonicity.} Because \(s'(p)=\tfrac{\pi}{2}\sin(\pi p)>0\) on $(0,1)$, \(s([0,1])=[0,1]\), and $\sin^2 x$ is strictly increasing on $(0,\tfrac{\pi}{2})$, the inner argument $\tfrac{\pi}{2}s(p)\in[0,\tfrac{\pi}{2}]$ is non-constant, so $g_{\text{alt}}$ is strictly increasing on $[0,1]$ and hits the boundaries. \textit{Monotonicity of $g_1$.} Since $g_1'(p)=6p-6p^2=6p(1-p)\ge 0$ on $[0,1]$ (strictly $>0$ on $(0,1)$), $g_1$ is strictly increasing and hits the boundaries $g_1(0)=0$, $g_1(1)=1$. Both examples satisfy Assumption~1 and reach $g(0)=0,\ g(1)=1$; they genuinely differ from $g(p)=\sin^2(\tfrac{\pi p}{2})$. (Note that $\tfrac{1-\cos(\pi p)}{2}\equiv \sin^2(\tfrac{\pi p}{2})$, so that cosine form is \emph{not} independent.)

\begin{figure}[h]
\centering
\includegraphics[width=0.7\textwidth]{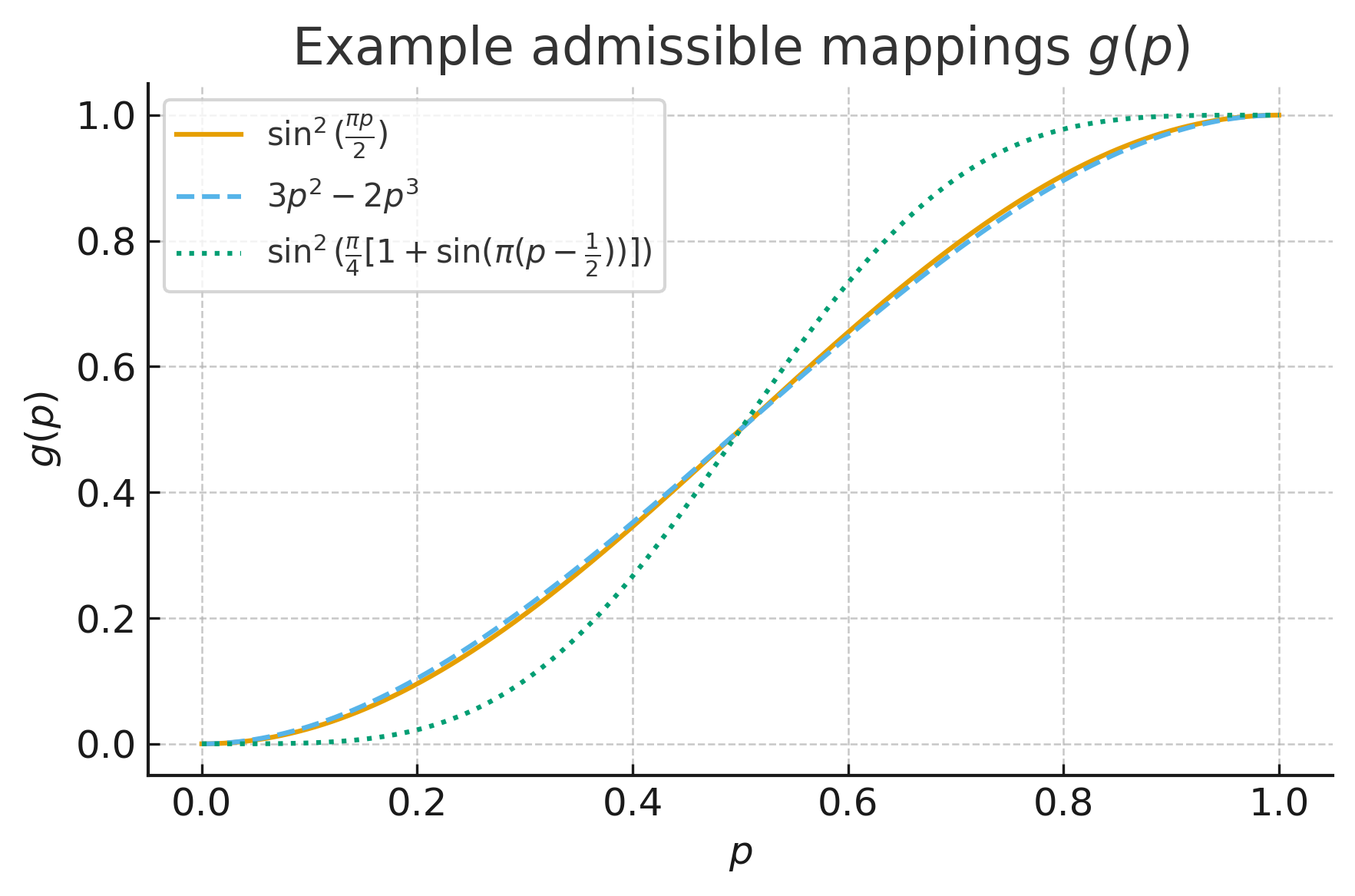} % or g_functions.pdf
\caption{Example mappings $g(p)$ satisfying Assumption~1 (complement preservation), continuity, and strict monotonicity.
Czachor’s choice $g(p)=\sin^2\!\big(\frac{\pi p}{2}\big)$ (solid) is one among many admissible forms, including
the polynomial $3p^2-2p^3$ (dashed) and a distinct, symmetric sine-embedded variant
$g_{\text{alt}}(p)=\sin^{2}\!\big(\frac{\pi}{4}[1+\sin(\pi(p-\frac{1}{2}))]\big)$ (dotted).
All three interpolate $(0,0)$ to $(1,1)$.}
\end{figure}

\subsection*{Step 5 — About Bell’s theorem}
On the hidden-variable side we now operate with a non-Kolmogorov probability calculus, so standard Bell-inequality derivations don’t go through as usual.\footnote{Equivalent formulations of Bell’s locality condition include \emph{factorizability}, i.e.\ the \emph{conjunction} of outcome independence and parameter independence, within the classical (Kolmogorov) framework. We use the common shorthand of locality, realism, and independence of settings.}
On the observable side we have chosen \(g\) to match quantum statistics.

That said, matching the full singlet correlation \(E(\varphi)=-\cos\varphi\) (and CHSH (Clauser--Horne--Shimony--Holt) violations) requires a \emph{joint} model—rules for how settings and hidden variables \(\lambda\) combine. Here $\varphi$ denotes the relative angle between the measurement axes at the two sites for the spin-singlet state. The single-argument map \(g\) fixes only marginals; without an explicit joint (how \(a,b,\lambda\) combine), \(E(\varphi)\) is underdetermined.

\subsection*{Step 6 — Where the frameworks get mixed}
The abstract hierarchy (Step 1) is independent of Kolmogorov assumptions. But the specific \(g\) used to match QM (Step 4) is \emph{calibrated by} Born’s rule and FS geometry—i.e., by standard quantum structure at the observable level (Kolmogorov within each fixed context).
So the construction simultaneously changes the hidden-variable calculus while importing a quantum-calibrated \(g\) at the top. That’s the mixing.

\subsection*{Step 7 — What follows from this}
If one truly abandons observable-level Kolmogorov structure, there is no special reason to prefer \(g(p)=\sin^2(\tfrac{\pi p}{2})\); Assumption 1 leaves infinitely many choices. The close fit to quantum correlations is therefore largely \emph{by construction}, not a new physical derivation.

In that sense, the proposal is not a counterexample to Bell’s theorem in its original scope. It modifies a standard premise (classical Kolmogorov probability) and then reuses a quantum-calibrated \(g\), so agreement with QM is built in rather than emergent. % (In one precise formulation, Bell’s theorem follows from classical Kolmogorov probability with \emph{factorizability} and \emph{measurement independence}.)
 
\section*{Conclusion}

Czachor's proposal starts from a formally coherent observation about isomorphic arithmetic hierarchies, but the specific framework he constructs does not hold up under closer scrutiny. Assumption~1, while central to the argument, leaves the mapping $g$ largely underdetermined—yet Czachor proceeds to anchor it using the Born rule and the Fubini--Study metric, both of which originate in standard quantum theory where Born probabilities are Kolmogorov \emph{within a fixed measurement context}. This step, crucially, reintroduces the very arithmetic formalism that the hierarchy model was intended to move beyond. What emerges is not a clean break from classical probability, but rather a \emph{hybrid Kolmogorov/non-Kolmogorov model}—an uneasy blend wherein the core mechanism relies on principles it ostensibly rejects.

Within this setup, the emergence of quantum singlet correlations isn't a surprising result, but an engineered outcome—an artefact of how $g$ is chosen. Because this choice is made rather than derived, the model fails to qualify as a counterexample to Bell's theorem. Since it modifies a standard probability premise in the theorem’s usual (Kolmogorov-based) derivations, it \emph{falls outside} that scope; the approach reframes the premises rather than refuting them. The apparent reproduction of quantum correlations only shows that, by redefining the underlying calculus, one can reassemble results that Bell showed to be unattainable within classical constraints. In that regard, Czachor’s construction serves more as a reinterpretation of how probabilities are modeled than a disproof of Bell’s logic.

Despite this, Czachor has presented his framework as something far more transformative. In a public interview \cite{Czachor_Inter} he stated:

\begin{quote}
“\ldots I am changing the paradigm of physics. I propose that, just as Einstein incorporated geometry into physics so that geometry ceased to be something abstract, Kantian and \textit{a priori}, and instead became a branch of physics — which is the essence of general relativity — at this moment I claim that an analogous status applies to arithmetic structures. They should not be treated as something given \textit{a priori}, but as something to be used in the most general way possible, and then let experiment decide what is concretely valid. In this sense it is a new paradigm, and theorists of science say that a revolution takes place when the paradigm changes. Such a paradigm shift was relativity theory \ldots I am incorporating arithmetic into physics \ldots perhaps nature adds and multiplies differently than we imagine \ldots"
\end{quote}

The analogy to Einstein’s revolution is certainly evocative—but it oversells the current state of the formalism. Our examination of Assumption~1 in \cite{Czachor_Hierarchy} reveals that $g(p)$ is far from uniquely determined. An infinite family of admissible functions fits the loose constraints, and Czachor’s sinusoidal choice is lifted wholesale from conventional quantum mechanics. The framework thereby ends up recycling elements it initially set out to replace.

A parallel issue appears across Czachor’s broader body of work on generalized calculus \cite{CommentCzachor11, arxiv2508}, where different bijections $f$ can produce either physically consistent outcomes (such as valid relativistic velocity addition via $f(\beta) = \artanh(\beta)$) or unphysical results. In general, the induced operation $\beta_1\oplus\beta_2=f^{-1}\!\big(f(\beta_1)+f(\beta_2)\big)$ is only a \emph{partial} operation on $(-1,1)$ unless $\operatorname{Im}(f)$ is closed under addition (e.g., $f=\artanh$ has $\operatorname{Im}(f)=\mathbb{R}$). For instance, taking $\beta_1=\beta_2=0.9$ with $f(\beta)=\beta^3$ gives $f(\beta_i)=0.729$ and $f(\beta_1)+f(\beta_2)=1.458$, which lies \emph{outside} $\operatorname{Im}(f)=(-1,1)$. Hence $\beta_1\oplus\beta_2$ is \emph{not defined} without extending $f^{-1}$ beyond its natural domain. If one \emph{does} extend $f^{-1}$ to $\mathbb{R}$, one obtains $(1.458)^{1/3}\approx 1.134$, which merely signals \emph{loss of closure} of the operation rather than a physical “superluminal sum.” The fact that both physically acceptable and unacceptable cases emerge from the same formal approach highlights the absence of internal selection criteria—a significant limitation when claiming to offer a foundational overhaul of physics.

Taken together, these issues make Czachor’s rhetoric about a paradigm shift seem premature. Unlike general relativity, which addressed empirical anomalies and led to testable predictions, the hierarchical arithmetic framework has yet to yield new insights or experimental implications. Instead, it repackages known results under a new formal guise. The aspirational tone of Ref.~\cite{Czachor_Hierarchy}, particularly in \textit{Sec. 14} with its closing line—``\textit{A new scientific paradigm is on the horizon.}''—currently reads more like philosophical ambition than a scientific breakthrough. See also \cite{arxiv2508, Sienicki_Sienicki_2025}.

\end{document}